\documentclass[twocolumn]{aastex62}
\usepackage{natbib}
\usepackage{epsfig}
\usepackage{graphicx}
\usepackage{subfigure}
\usepackage{float}
\usepackage{amsmath}
\usepackage{color}
\usepackage{amssymb}
\usepackage{amsfonts}
\usepackage{apjfonts}
\newcommand{\PMO}{Purple Mountain Observatory, Chinese Academy of Sciences, Nanjing 210023, China; jjwei@pmo.ac.cn, xfwu@pmo.ac.cn}
\newcommand{\USTC}{School of Astronomy and Space Sciences, University of Science and Technology of China, Hefei 230026, China; daizg@ustc.edu.cn}
\newcommand{\NJU}{School of Astronomy and Space Science, Nanjing University, Nanjing 210023, China}
\newcommand{\NAOC}{CAS Key Laboratory of FAST, NAOC, Chinese Academy of Sciences, Beijing 100101, China}
\newcommand{\UCAS}{University of Chinese Academy of Sciences, Beijing 100049, China}
\newcommand{\UNLV}{Department of Physics and Astronomy, University of Nevada, Las Vegas, Las Vegas, NV 89154, USA}
\newcommand{\UKZN}{NAOC-UKZN Computational Astrophysics Centre, University of KwaZulu-Natal, Durban 4000, South Africa}

\shortauthors{Wei et al.}

\begin{document}

\title{Similar Scale-invariant Behaviors between Soft Gamma-ray Repeaters and An Extreme Epoch from FRB 121102}

%\correspondingauthor{Jun-Jie Wei}
%\email{jjwei@pmo.ac.cn}

\author{Jun-Jie Wei}
\affiliation{\PMO}
\affiliation{\USTC}

\author{Xue-Feng Wu}
\affiliation{\PMO}
\affiliation{\USTC}

\author{Zi-Gao Dai}
\affiliation{\USTC}
\affiliation{\NJU}

\author{Fa-Yin Wang}
\affiliation{\NJU}

\author{Pei Wang}
\affiliation{\NAOC}

\author{Di Li}
\affiliation{\NAOC}
\affiliation{\UCAS}
\affiliation{\UKZN}

\author{Bing Zhang}
\affiliation{\UNLV}

\begin{abstract}
The recent discovery of a Galactic fast radio burst (FRB) associated with a hard X-ray burst from the
soft gamma-ray repeater (SGR) J1935+2154 has established the magnetar origin of at least some FRBs.
In this work, we study the statistical properties of soft gamma-/hard X-ray bursts from
SGRs 1806--20 and J1935+2154 and of radio bursts from the repeating FRB 121102. For SGRs,
we show that the probability density functions for the differences of fluences, fluxes, and durations at different
times have fat tails with a $q$-Gaussian form. The $q$ values in the $q$-Gaussian distributions are approximately
steady and independent of the temporal interval scale adopted, implying a scale-invariant structure of SGRs.
These features indicate that SGR bursts may be governed by a self-organizing criticality (SOC) process,
confirming previous findings. Very recently, 1652 independent bursts from FRB 121102 have been
detected by the Five-hundred-meter Aperture Spherical radio Telescope (FAST). Here we also investigate the
scale-invariant structure of FRB 121102 based on the latest observations of FAST, and show that FRB 121102
and SGRs share similar statistical properties.
Given the bimodal energy distribution of FRB 121102 bursts, we separately explore the scale-invariant behaviors
of low- and high-energy bursts of FRB 121102. We find that the $q$ values of low- and high-energy bursts are different,
which further strengthens the evidence of the bimodality of the energy distribution.
Scale invariance in both the high-energy component of FRB 121102 and SGRs can be well explained within
the same physical framework of fractal-diffusive SOC systems.
\end{abstract}

\keywords{Magnetars --- Soft gamma-ray repeaters --- Radio transient sources --- Radio bursts}

\section{Introduction}
%slowly driven nonlinear dissipative systems
Self-organized criticality (SOC; \citealt{1986JGR....9110412K,1987PhRvL..59..381B,2016SSRv..198...47A})
has been found in dynamical behaviors of astrophysics systems. We here examine two types of transient phenomena,
namely, soft gamma-ray repeaters (SGRs) and fast radio bursts (FRBs), which have been shown to be associated with
each other in at least one case \citep{2020Natur.587...45Z}.

Magnetars are highly magnetized neutron stars that exhibit dramatic variability over a broad range of
timescales (\citealt{1992ApJ...392L...9D,1998Natur.393..235K}, see also \citealt{2015RPPh...78k6901T,2017ARA&A..55..261K}
for reviews). Many magnetars were first observed as SGRs, i.e., sources of repeated bursts of soft gamma-/hard X-rays, with
typical durations of $\sim0.1-1$ s and peak luminosities of $\sim10^{39}-10^{41}$ erg $\rm s^{-1}$.
Three several-minute-long giant flares with peak luminosities up to $\sim10^{44}-10^{47}$ erg $\rm s^{-1}$
have also been observed from three different SGRs \citep{1979Natur.282..587M,1999Natur.397...41H,2005Natur.434.1107P}.
Both bursts and flares are believed to be powered by the dissipation and decay of the ultrastrong
magnetic fields, either through neutron star crustquakes \citep{1995MNRAS.275..255T} or magnetic reconnection
\citep{2003MNRAS.346..540L}.

Several works have found that the energy distribution of SGR bursts can be well fitted by a power-law function
(e.g., \citealt{1996Natur.382..518C,1999ApJ...526L..93G,2000ApJ...532L.121G,2012ApJ...755....1P,2020MNRAS.491.1498C}).
Power-law size distributions of extreme events can be explained by the concept of SOC in slowly driven nonlinear
dissipative systems \citep{1987PhRvL..59..381B,2016SSRv..198...47A}. The SOC subsystems will self-organize,
owing to some driving force, to a critical state, at which a small local disturbance can produce an avalanche-like
chain reaction of any size within the system \citep{1987PhRvL..59..381B}. The fundamental property of all
SOC systems have in common is the emergence of scale-free power-law size distributions
\citep{2011soca.book.....A,2012A&A...539A...2A,2015ApJ...814...19A,2013NatPh...9..465W,2020FrPhy..1614501L,2021FrPhy..1624503L}.
Given the earthquake-like power-law energy distributions, \cite{1999ApJ...526L..93G} suggested,
for the first time, that systems responsible for SGR bursts are in a SOC state.
The earthquake-like behavior also suggests that the energies of SGRs originate from starquakes of magnetars \citep{1992ApJ...392L...9D,1995MNRAS.275..255T}.

Besides the power-law size distribution, another hallmark of SOC systems is the scale invariance of the avalanche
size differences \citep{2007PhRvE..75e5101C,2015EPJB...88..206W}. \cite{2007PhRvE..75e5101C} showed that
the probability density functions (PDFs) for the earthquake energy differences at different times have
fat tails with a $q$-Gaussian form, which could be well explained by introducing a small-world topology on
the Olami-Feder-Christensen (OFC) model \citep{1992PhRvL..68.1244O}. In physics, the OFC model is one of
the most popular models displaying SOC. The $q$-Gaussian distribution was then suggested as an important
character for describing the presence of criticality \citep{2007PhRvE..75e5101C}. Moreover, the form of the PDF
does not depend on the time interval adopted for the earthquake energy difference, i.e., the $q$ values
in $q$-Gaussian distributions keep nearly constant for different scale intervals, which indicates that
there is a scale-invariant structure in the energy differences of earthquakes \citep{2007PhRvE..75e5101C,2015EPJB...88..206W}.
Subsequently, \cite{2017ChPhC..41f5104C} studied 384 X-ray bursts observed in three active episodes of SGR J1550--5418
and found that this SGR shows similar scale-invariant behaviors, i.e., the PDFs of the
differences of fluences, peak fluxes, and durations exhibit a common $q$-Gaussian distribution at different
scale intervals.

FRBs are intense millisecond-duration bursts of radio waves occurring in the universe
\citep{2007Sci...318..777L,2019A&ARv..27....4P,2019ARA&A..57..417C}. So far, more than 600 FRBs
have been reported, and over two dozens of them have been seen to repeat (e.g.,
\citealt{2016Natur.531..202S,2019ApJ...885L..24C,2020ApJ...891L...6F,2021arXiv210604352T}). Although rapid developments have
been made in the FRB research field, the physical origin of FRBs remains mysterious
\citep{2019PhR...821....1P,2020Natur.587...45Z,2021arXiv210304165G,2021SCPMA..6449501X}. Magnetars have been proposed as the
likely engine to power repeating FRBs \citep{2010vaoa.conf..129P,2014ApJ...797...70K,2016MNRAS.461.1498M,
2016ApJ...826..226K,2017ApJ...841...14M,2017JCAP...03..023W,2017ApJ...843L..26B,2017MNRAS.468.2726K,
2018ApJ...868...31Y,2019ApJ...879....4W,2020MNRAS.491.1498C,2020ApJ...891...72W}. On 2020 April 28, one FRB-like event was
independently detected by the Canadian Hydrogen Intensity Mapping Experiment \citep{2020Natur.587...54C}
and the Survey for Transient Astronomical Radio Emission 2 \citep{2020Natur.587...59B} in association with
a hard X-ray burst from the Galactic magnetar SGR J1935+2154 during its active phase \citep{2021NatAs.tmp...54L,
2020ApJ...898L..29M,2021NatAs...5..372R,2021NatAs...5..401T}. This discovery strongly supports that magnetar
engines can produce at least some extragalactic FRBs.

Recently, by analyzing 93 bursts from the repeating FRB 121102 by the observation of the Green Bank
Telescope at 4--8 GHz \citep{2018ApJ...866..149Z}, \cite{2020MNRAS.491.2156L} found that FRB 121102 has
the property of scale invariance similar to that of SGR J1550--5418. During an extreme episode of bursts,
FRB 121102 produced 1652 detectable events within a time-span of 62 days \citep{2021arXiv210708205L}.
The peak burst rate reached 117 hr$^{-1}$. Such a range of cadence facilitate further investigation into
the SOC structures of this source. Another key feature of this burst set from FRB 121102 is the bimodal
energy distribution of the burst rate.

Among the 16 known SGRs (12 confirmed and 4 candidates; \citealt{2014ApJS..212....6O}), SGR J1550--5418 is
the only source to date that has been used to investigate the scale-invariant property \citep{2017ChPhC..41f5104C}.
So it is interesting to know whether other SGRs (e.g., SGR 1806--20 and SGR J1935+2154) share the same
property. The magnetar research group at Sabanc{\i} University presented their systematic temporal and broad-band
spectral analysis of over 1,500 bursts from SGR J1550--5418, SGR 1900+14, and SGR 1806--20 observed with
the Rossi X-ray Timing Explorer (RXTE; e.g., \citealt{2017ApJS..232...17K}). In these public data, SGR 1806--20
has the largest burst sample (924 bursts), which will be used in our following analysis. Additionally, since
SGR J1935+2154 is the first source that has been reported to power a Galactic FRB \citep{2020Natur.587...54C,
2020Natur.587...59B}, we will also investigate the physical connection between SGR J1935+2154 and repeating
FRBs.

In this paper, we investigate the scale-invariant behaviors of SGR 1806--20, SGR J1935+2154, and FRB 121102.
For SGRs, the PDFs of the differences of fluences (or total counts), peak fluxes (or peak counts), and
durations at different time scales are shown in Section~\ref{sec:PDFs}. These PDFs can be well fitted by
a $q$-Gaussian function, and the functional form does not depend on the scale interval adopted.
In Section~\ref{sec:compare}, we compare the PDFs between SGRs and repeating FRB 121102. The evidence that
may shed light on the bimodal burst energy distribution of FRB 121102 is discussed in Section~\ref{sec:evidence}.
Lastly, a brief summary and discussion are given in Section~\ref{sec:summary}.

%Magnetars are a promising candidate for the origin of Fast Radio Bursts (FRBs).

\section{Scale-invariance in SGRs}
\label{sec:PDFs}

\subsection{Data}
The magnetar research group at Sabanc{\i} University has constructed an online database of magnetar bursts
detected by the RXTE between 1996 and 2011, which can be visited at \href{http://magnetars.sabanciuniv.edu}{http://magnetars.sabanciuniv.edu}.
This database contains 924 bursts from SGR 1806--20, representing the largest sample for a single observation.
The trigger time, total counts, peak counts, and duration of each burst are available in the database.

For SGR J1935+2154, the bursts observed by the Gamma-ray Burst Monitor (GBM) onboard the Fermi satellite during
the source's six active episodes from 2014 to 2020 are used \citep{2020ApJ...893..156L,2020ApJ...902L..43L}.
The number of bursts is 260. The available information for each burst include trigger time, fluence, and duration,
but not peak flux.

The 924 bursts from SGR 1806--20 were all observed with the RXTE/Proportional Counter Array (PCA), and
the 260 bursts from SGR J1935+2154 were all observed with the Fermi/GBM. These two instruments are different
in numerous ways, including energy passband, sensitivity, background level, etc. The comparison of total counts detected with
a pointing instrument, RXTE/PCA (effectively sensitive to 3--30 keV photons) to fluences of an all-sky monitor,
GBM (8--200 keV) has to be handled with care. A conversion factor between each PCA count and GBM fluence,
in principle, has to be determined \citep{1999ApJ...526L..93G,2000ApJ...532L.121G,2001ApJ...558..228G}.
On the other hand, since we are interest in the statistical distribution of the total count differences at different
time intervals (see more below), the difference between two PAC counts may be approximated as the difference
between two GBM fluences.

The adopted durations of all bursts from both SGR 1806--20 and SGR J1935+2154 were determined by
a Bayesian block algorithm, but the sensitivity of detecting instruments are expected to affect the resulting
duration distributions. Thus, there is a caveat that the comparison of the PDFs of the duration differences between
SGR 1806--20 and SGR J1935+2154 may be affected by the sensitivity of different instruments. Note that
\cite{2017ChPhC..41f5104C} studied the statistical properties of a sample of 384 SGR J1550--5418 bursts
detected with the Fermi/GBM. With the same detecting instrument, the duration comparison between SGR J1935+2154
and SGR J1550--5418 would be credible.
Here we estimate the waiting time between two adjacent bursts through
$\Delta t=t_{i+1}-t_{i}$, where $t_{i+1}$ and $t_{i}$ are the trigger times of the $(i+1)$-th and $i$-th bursts,
respectively. As both SGR 1806--20 and SGR J1935+2154 consist of several isolated epochs, we discard the waiting
time between the last burst of a certain epoch and the first burst of the next epoch.

\begin{table*}
\centering \caption{The mean values of $q$ in the $q$-Gaussian distribution for earthquakes, SGRs, and FRB 121102}
\begin{tabular}{lcccccc}
\hline
\hline
 Phenomena &  Energy range (erg)   &  Number  &  $q$-fluence (energy) & $q$-flux & $q$-duration & References  \\
\hline
earthquakes      &     &  400,000 & $1.75\pm0.15$  & -- & --  &  \cite{2007PhRvE..75e5101C}\\
\hline
SGR J1550--5418  &     &  384  &   $2.41\pm0.29$  & $2.40\pm0.30$ & $2.06\pm0.23$   & \cite{2017ChPhC..41f5104C}  \\
SGR 1806--20     &     &  924  &   $2.72\pm0.05$  & $2.65\pm0.05$ &  $2.48\pm0.05$  & This work  \\
SGR J1935+2154   &     &  260  &   $2.78\pm0.12$  & -- & $2.28\pm0.15$              & This work \\
\hline
FRB 121102       & Overall bursts  &  1652 &   $2.51\pm0.03$  & $2.50\pm0.03$ & $1.42\pm0.04$   & This work \\
                 & with $4\times10^{36}<E<8\times10^{39}$   &  &     &  &    &  \\
\hline
FRB 121102       & Low-energy bursts &  1253 &   $1.63\pm0.09$  & $1.81\pm0.05$ & $1.39\pm0.06$   & This work \\
                 & with $4\times10^{36}<E\leq2\times10^{38}$ &   &    &  &    &  \\
                & High-energy bursts &  399 &   $2.21\pm0.10$  & $2.34\pm0.09$ &$1.50\pm0.16$   & This work \\
                & with $2\times10^{38}<E<8\times10^{39}$  &   &  &  &   & \\
\hline
\end{tabular}
\label{table1}
\end{table*}

\begin{figure*}
\begin{center}
\vskip-0.1in
\includegraphics[width=0.45\textwidth]{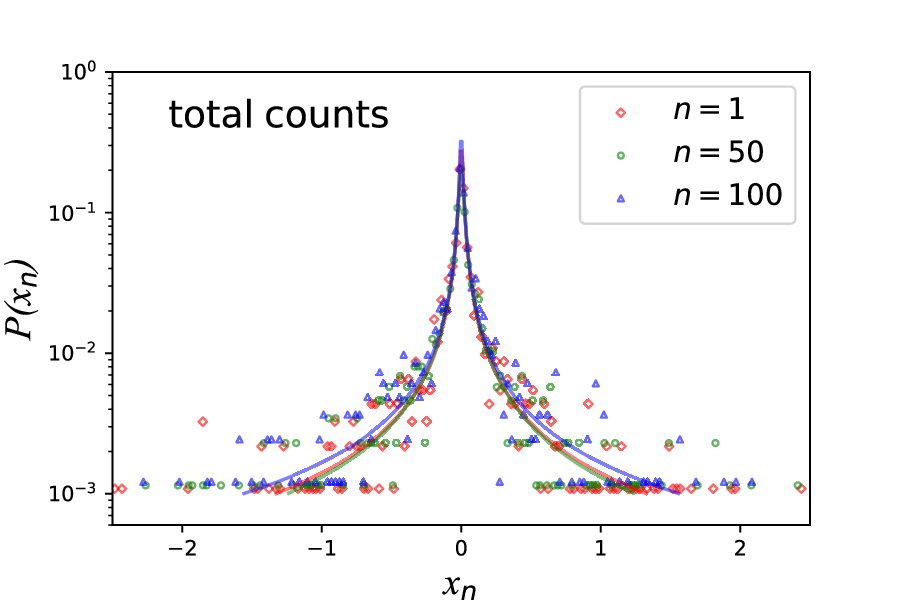}
\includegraphics[width=0.45\textwidth]{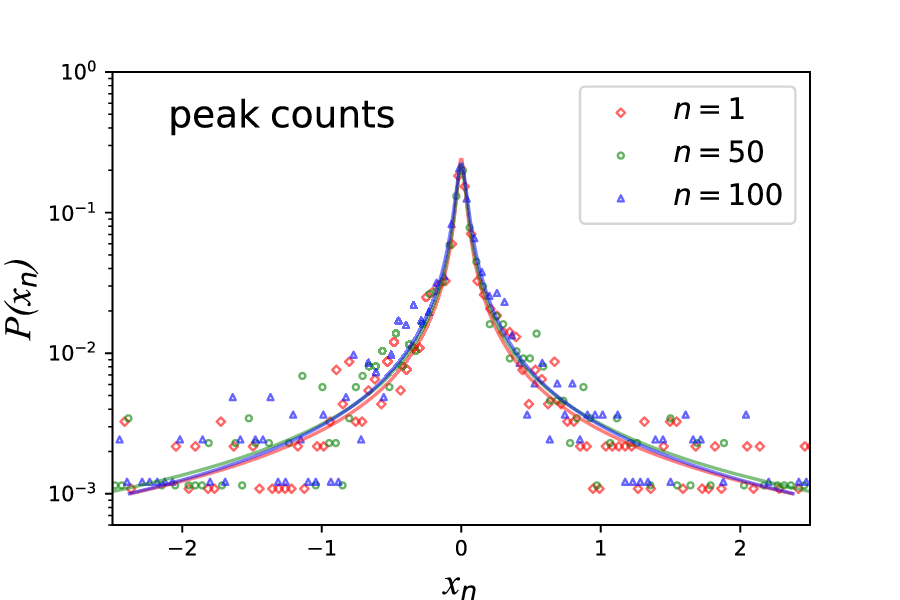}
\includegraphics[width=0.45\textwidth]{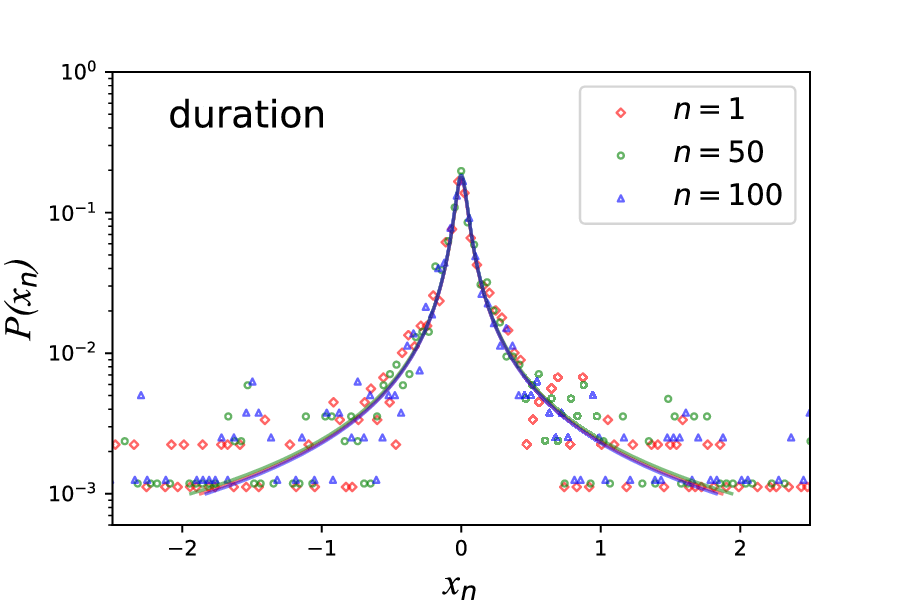}
\includegraphics[width=0.45\textwidth]{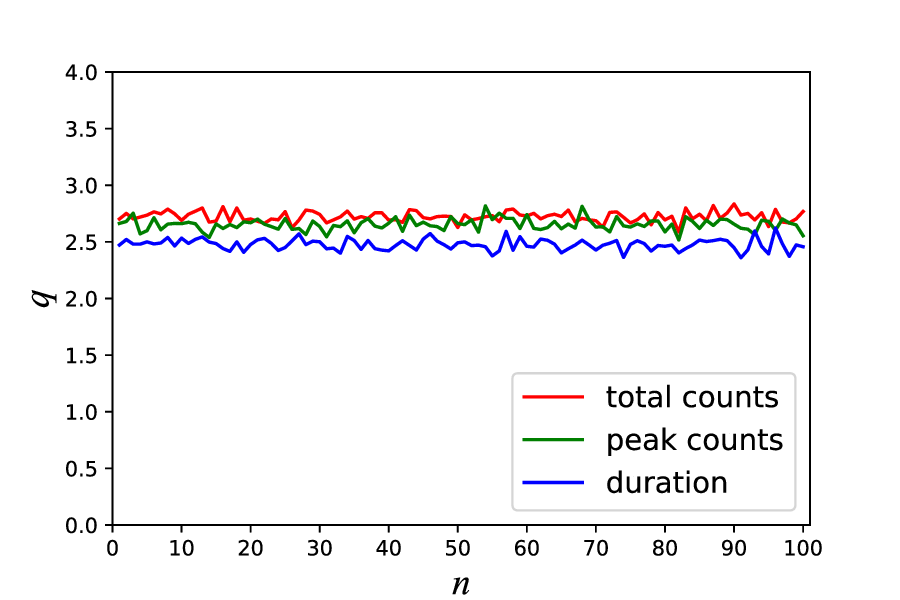}
%\vskip-0.1in
\caption{Scale-invariant structure of SGR 1806--20. PDFs of the differences of total counts
(upper-left panel), peak counts (upper-right panel), and durations (lower-left panel) for $n=1$ (red diamonds),
$n=50$ (green circles), and $n=100$ (blue triangles), and the best-fitting $q$-Gaussian distributions (solid lines).
Lower-right panel: the best-fitting $q$ values in the $q$-Gaussian distribution as a function of $n$.}
\label{fig:f1}
\vskip-0.2in
\end{center}
\end{figure*}

\subsection{PDFs of the Avalanche Size Differences}
We are now in position to use the data of SGR 1806--20 and SGR J1935+2154 to investigate the PDFs
of the avalanche size differences. The difference between two avalanche sizes is given by
\begin{equation}
X_{n}=S_{i+n}-S_{i}\;,
\end{equation}
where $S_{i}$ is the size (fluence, peak flux, duration, or waiting time) of the $i$-th burst
in temporal order and $n$ (an integer) is the temporal interval scale. In practice, $X_{n}$ is
normalized to
\begin{equation}
x_{n}=\frac{X_{n}}{\sigma_{X_{n}}}\;,
\end{equation}
where $\sigma_{X_{n}}$ is the standard deviation of $X_{n}$. We are interest in
the statistical distribution of $x_{n}$.

Figure~\ref{fig:f1} shows the statistical results of SGR 1806--20. In this plot, we display
the PDFs of the differences of total counts (upper-left panel), peak counts (upper-right panel),
and durations (lower-left panel) for $n=1$ (diamonds), $n=50$ (circles), and $n=100$ (triangles).
Following \cite{2017ChPhC..41f5104C}, the data are binned based on the Freedman-Diaconis rule
\citep{Freedman1981}. One can see from Figure~\ref{fig:f1} that these PDFs $P(x_{n})$ have
a sharp peak and fat tails, which are different from Gaussian behaviors. The sharp peak
means that small size differences are most likely to happen, while the fat tails suggest that
there are rare but relatively large size differences. The large fluctuations presented in the tails
are caused by the incompleteness given from the lack sampling of small magnitude events at the global
scale \citep{2007PhRvE..75e5101C}. Additionally, the data points in Figure~\ref{fig:f1}
are almost independent of the temporal interval scale $n$ adopted for the avalanche size
difference, indicating a common form of $P(x_{n})$. Here we use the Tsallis $q$-Gaussian function
\citep{1988JSP....52..479T,1998PhyA..261..534T}
\begin{equation}
f(x_{n})=\alpha\left[1-\left(1-q\right)x_{n}^{2}/\beta\right]^{\frac{1}{1-q}}
\label{eq:q-Gauss}
\end{equation}
to fit $P(x_{n})$, where $\alpha$, $\beta$, and $q$ are free parameters. When $q\rightarrow1$,
the $q$-Gaussian distribution reduces to the Gaussian distribution with mean zero and standard deviation
$\sigma=\sqrt{\beta/2}$. Thus, $q\neq1$ denotes a departure from Gaussian statistics.

The best-fitting parameters ($\alpha$, $\beta$, and $q$) are obtained by minimizing the $\chi^{2}$ statistics,
\begin{equation}
\chi^{2}=\sum_{i}\frac{\left[f(x_{n,i})-P(x_{n,i})\right]^{2}}{\sigma_{P,i}^{2}}\;,
\end{equation}
where $\sigma_{P,i}=\sqrt{N_{{\rm bin},i}}/N_{\rm tot}$ is the uncertainty of the data
point,\footnote{For clarity, the uncertainties of the data points are not presented in the figure.}
with $N_{{\rm bin},i}$ is the event number in the $i$-th bin and $N_{\rm tot}$ is the total number of $x_{n}$.
We use the python implementation, emcee \citep{2013PASP..125..306F}, to apply the Markov Chain Monte Carlo method
to derive the best-fitting values and their corresponding uncertainties for these parameters.
In Figure~\ref{fig:f1}, the red, green, and blue smooth curves correspond to the best-fitting results for
$n=1$, $n=50$, and $n=100$, respectively. We may see that the PDFs of the differences of total counts,
peak counts, and durations are well fitted by the $q$-Gaussian function, and that the three curves in each panel
overlap almost completely. However, the PDF of the waiting time differences can not be well fitted by $q$-Gaussian
(see also \citealt{2017ChPhC..41f5104C}), so we do not exhibit it in the figure. The waiting time shows different
behaviors, which may be due to the discontinuous observations of the telescope.

\begin{figure*}
\begin{center}
\vskip-0.1in
\includegraphics[width=0.33\textwidth]{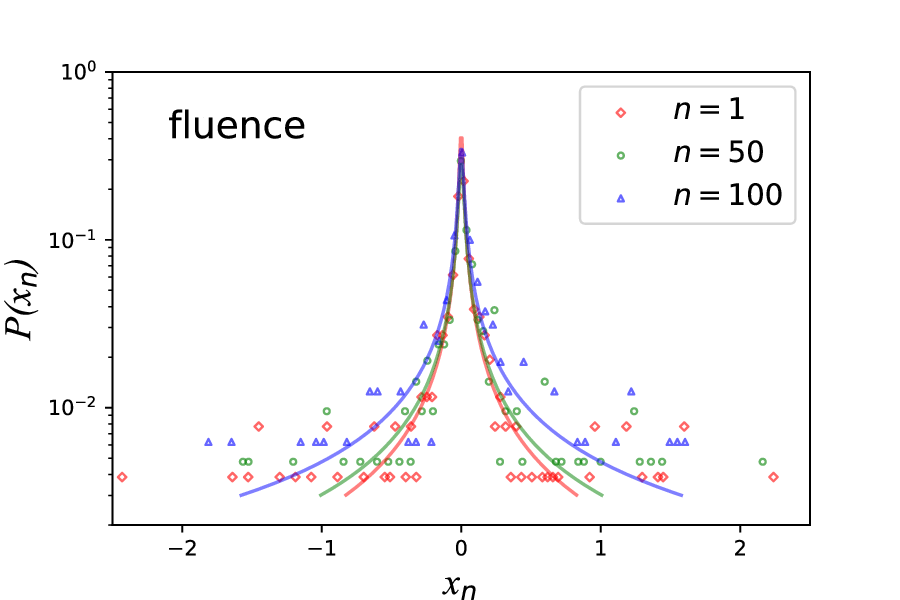}
\includegraphics[width=0.33\textwidth]{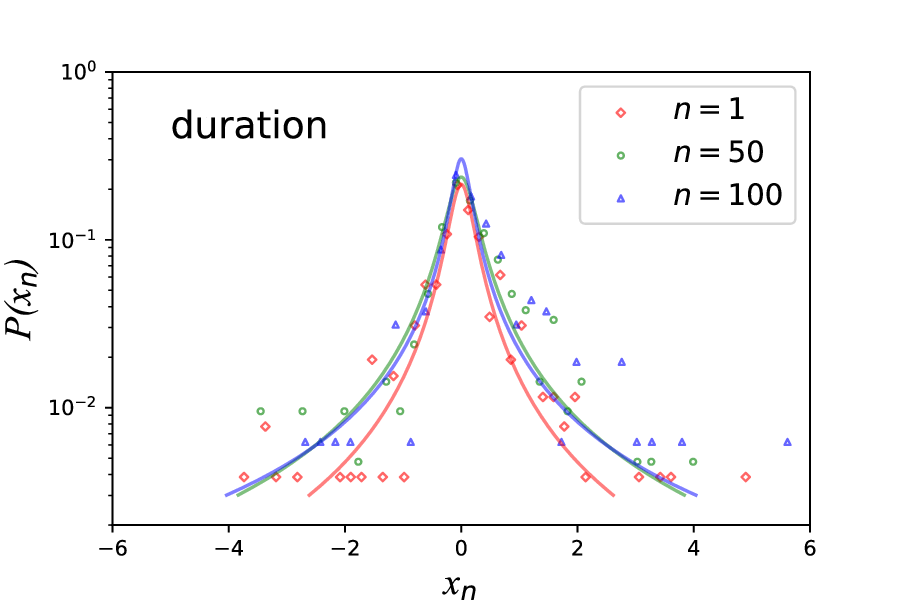}
\includegraphics[width=0.33\textwidth]{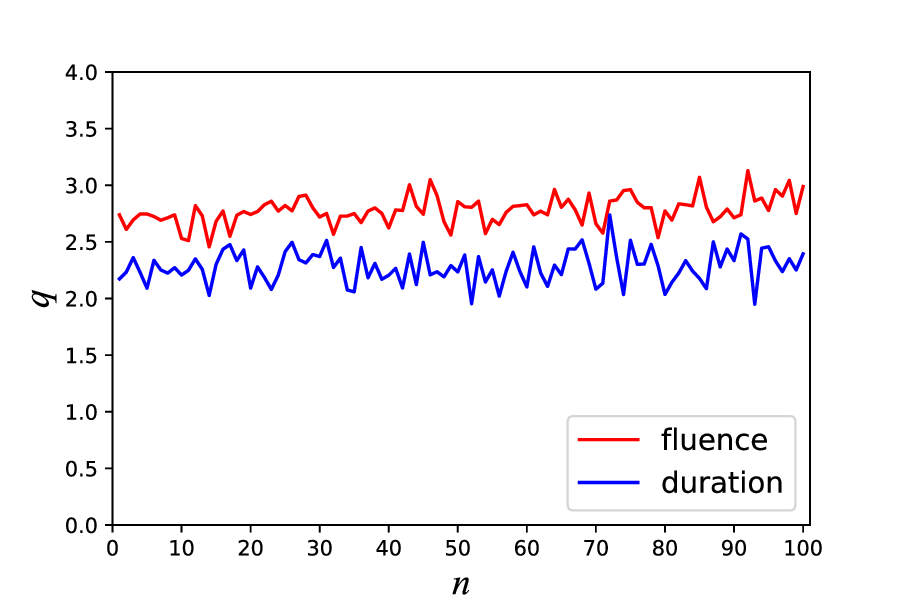}
%\vskip-0.1in
\caption{Scale-invariant structure of SGR J1935+2154. PDFs of the differences of fluences
(left panel) and durations (middle panel) for $n=1$ (red diamonds), $n=50$ (green circles), and
$n=100$ (blue triangles), and the best-fitting $q$-Gaussian distributions (solid lines).
Right panel: the best-fitting $q$ values in the $q$-Gaussian distribution as a function of $n$.}
\label{fig:f2}
\vskip-0.2in
\end{center}
\end{figure*}

Furthermore, we compute the PDFs of the differences of total counts, peak counts, and durations at different
scale intervals $1\leq n \leq100$, and fit the PDFs with the $q$-Gaussian function. In the lower-right panel
of Figure~\ref{fig:f1}, we plot the best-fitting $q$ values as a function of $n$. We find that the $q$ values
are approximately invariant for different scale intervals $n$. The property of the independence of $q$ values
on $n$ is referred to as scale invariance. The mean values of $q$ for total counts, peak counts, and duration
are $2.72\pm0.05$, $2.65\pm0.05$, and $2.48\pm0.05$, respectively, which are summarized in Table~\ref{table1}.
Here the uncertainty denotes the standard deviation of $q$. Interestingly, the $q$ values we find here are
close to the values derived from SGR J1550--5418 \citep{2017ChPhC..41f5104C}, which indicates that
there is a common scale-invariant property in SGRs.

Similar results can be obtained for the data of SGR J1935+2154, see Figure~\ref{fig:f2}. The $q$ values for this
data also keep approximately steady for different scale intervals $n$. The mean values of $q$ for fluence and duration
are $2.78\pm0.12$ and $2.28\pm0.14$, respectively. As shown in Table~\ref{table1}, these values are well consistent with
those of SGR J1550--5418 \citep{2017ChPhC..41f5104C} and SGR 1806--20, supporting a common scale-invariant structure of
SGRs again.

It is worth mentioning that our statistic relies on the size difference between two successive events.
However, both RXTE and Fermi are low orbit instruments and the sources of interest would frequently be blocked
by the Earth. It may happen that there are other bursts in between some of those events listed in the literature,
and not recorded because the sources were occulted. Moreover, there are also some unresolved weak bursts
in the data. We have proved that for the successive burst data, by changing the time interval $n$, or by reshuffling
the time series, no change in the PDFs of the size differences is observed. In other words, the scale-invariant behaviors
of SGRs are only weakly sensitive to the absence of the obscured bursts or the unresolved weak bursts.

\section{Comparison with FRB 121102}
\label{sec:compare}
Owing to the association between FRB 200428 and an X-ray burst from the Galactic magnetar SGR J1935+2154
\citep{2020Natur.587...54C,2020Natur.587...59B,2021NatAs.tmp...54L,2020ApJ...898L..29M,2021NatAs...5..372R,2021NatAs...5..401T},
it is natural to consider whether radio bursts of repeating FRBs have a similar scale-invariant behavior.
In this section, we compare the PDFs between SGRs and repeating FRBs.

Recently, \cite{2021arXiv210708205L} reported the detection of 1652 independent bursts in 59.5 hours spanning 62 days using
the Five-hundred-meter Aperture Spherical radio Telescope (FAST; \citealt{2011IJMPD..20..989N,2018IMMag..19..112L}) at 1.05--1.45 GHz.
Such a uniform sample prevent complex selection effects, which can be introduced by different instruments and different
frequency bands used in observations. Here we focus on the PDFs for the differences of energies, fluxes, and durations
at different time scales. For a fixed $n$, we fit the PDF with the $q$-Gaussian function and extract the best-fitting
$q$ value. Then we vary $n$ and derive the $q$ values as a function of $n$.

\begin{figure*}
\begin{center}
\vskip-0.1in
\includegraphics[width=0.45\textwidth]{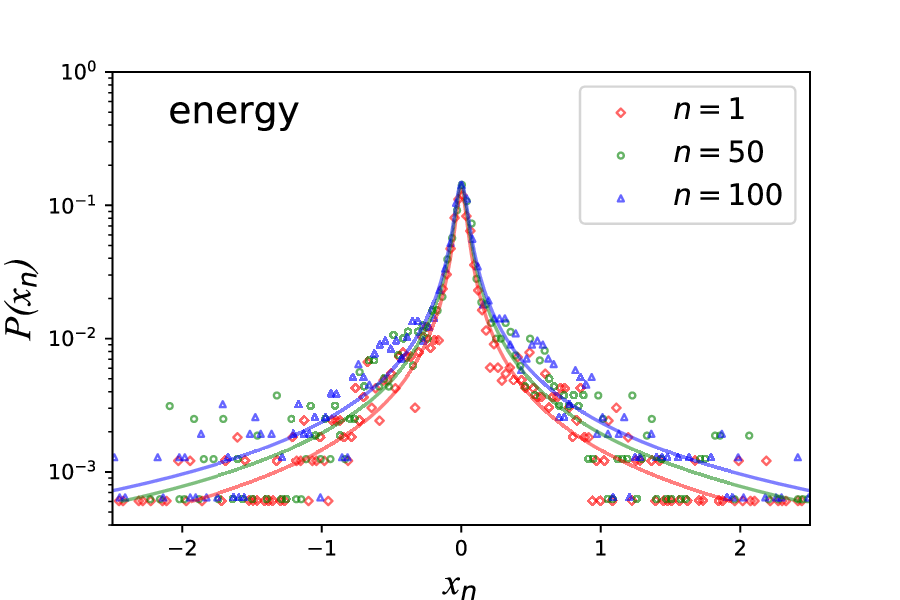}
\includegraphics[width=0.45\textwidth]{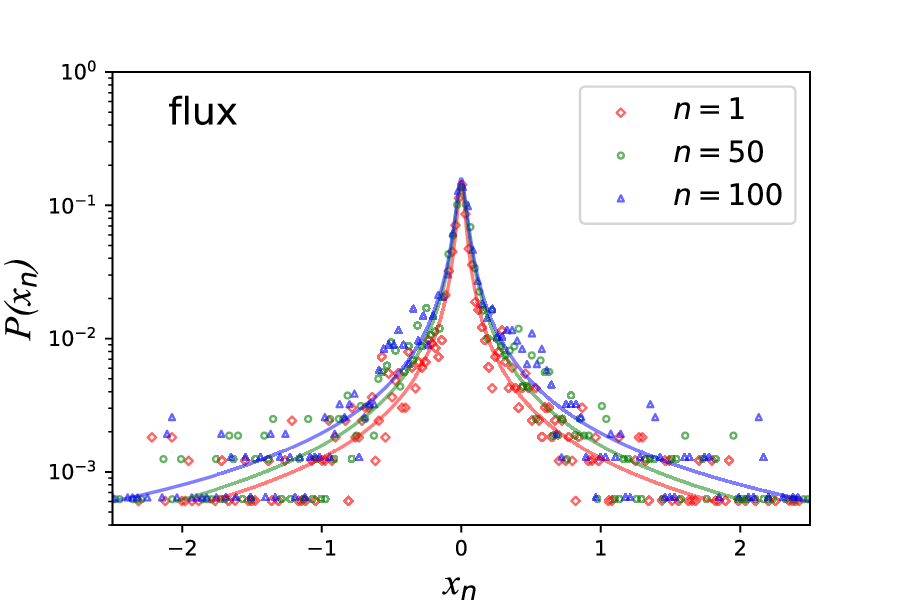}
\includegraphics[width=0.45\textwidth]{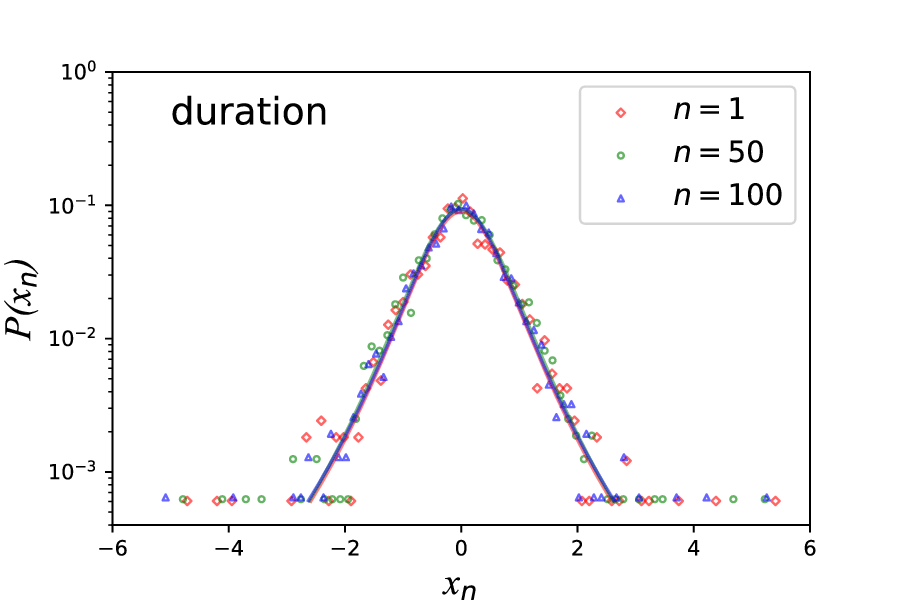}
\includegraphics[width=0.45\textwidth]{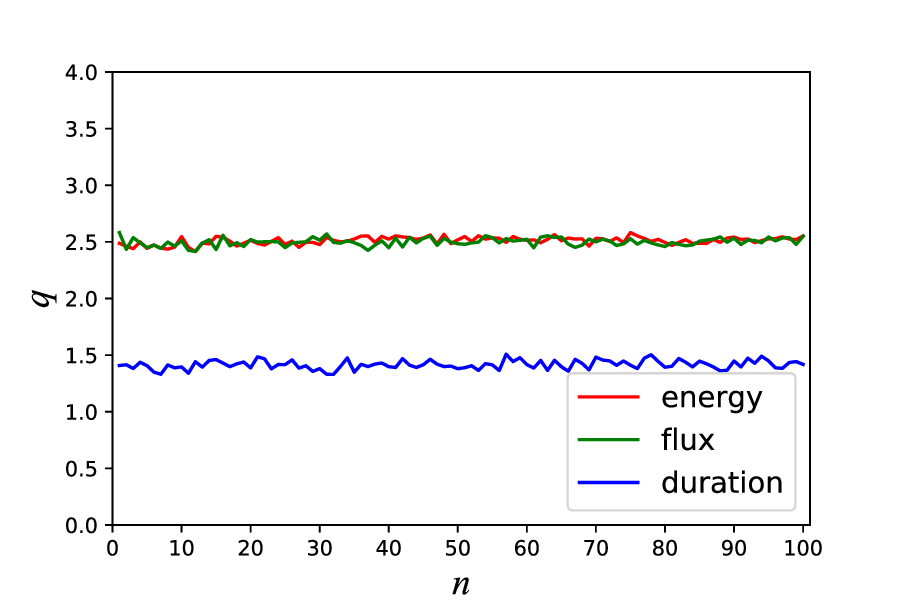}
%\vskip-0.1in
\caption{Scale-invariant structure of FRB 121102. PDFs of the differences of energies
(upper-left panel), fluxes (upper-right panel) and durations (lower-left panel) for $n=1$ (red diamonds),
$n=50$ (green circles), and $n=100$ (blue triangles), and the best-fitting $q$-Gaussian distributions (solid lines).
Lower-right panel: the best-fitting $q$ values in the $q$-Gaussian distribution as a function of $n$.}
\label{fig:f3}
\vskip-0.2in
\end{center}
\end{figure*}

Figure~\ref{fig:f3} exhibits some examples of the fits for the sample of FRB 121102. In this figure, we show
the PDFs of the differences of energies (upper-left panel), fluxes (upper-right panel),
and durations (lower-left panel) for $n=1$ (diamonds), $n=50$ (circles), and $n=100$ (triangles).
The solid curves stand for the best-fitting results. One can see that the PDFs can be well fitted by
the $q$-Gaussian function, in good agreement with the results of \cite{2020MNRAS.491.2156L}.
Similar to the scenario of SGRs, we also find that the $q$-Gaussian function gives a poor fit to the PDF of
the FRB waiting time differences, which is then not shown in the figure (see also \citealt{2020MNRAS.491.2156L}).
To better represent the scale-invariant property in FRB 121102, in the lower-right panel of Figure~\ref{fig:f3}
we plot the best-fitting $q$ values as a function of the temporal interval scale $n$. We confirm that
the $q$ values are almost stable and independent of $n$. This implies that there are similar scale-invariant
properties among earthquakes, SGRs, and FRBs.
We also list the average $q$ values of energy ($2.51\pm0.03$), flux ($2.50\pm0.03$), and duration ($1.42\pm0.04$)
for FRB 121102 in Table~\ref{table1}. We emphasize that the $q$-Gaussian distribution is a generalization of
the standard Gaussian distribution and it reduces to a Gaussian distribution when $q\rightarrow1$.
Since the average $q$ value of duration is relatively small (closing to $1$), the peak of the PDF in the lower-left
panel of Figure~\ref{fig:f3} is no longer very sharp as other PDFs.

\section{Scale-invariant Structures of the Bimodal Energy Distribution of FRB 121102 Bursts}
\label{sec:evidence}
The burst energy distribution of FRB 121102 can be adequately described as bimodal \citep{2021arXiv210708205L}.
At the low energy end ($\sim E<2\times10^{38}$ erg), a log-normal function can describe
the distribution reasonably well. While the distribution at the high energy end ($\sim E>2\times10^{38}$ erg)
can be well fitted by a generalized Cauchy function.
It is interesting to investigate whether low-energy bursts and high-energy bursts of FRB 121102 have a
similar scale-invariant behavior.

\subsection{Statistical properties of low- and high-energy bursts}

We divide the overall 1652 bursts into two components, i.e., 1253 low-energy bursts with $E\leq2\times10^{38}$
erg and 399 high-energy bursts with $E>2\times10^{38}$ erg. Then the scale-invariant structures of
the two components are investigated, respectively. As shown in Table~\ref{table1}, the average $q$ values
of energy, peak flux, and duration for the low-energy component are $q_{E}=1.63\pm0.09$, $q_{P}=1.81\pm0.05$,
and $q_{T}=1.39\pm0.06$. The average $q$ values for the high-energy component are $q_{E}=2.21\pm0.10$,
$q_{P}=2.34\pm0.09$, and $q_{T}=1.50\pm0.16$.
Interestingly, the average $q$ value obtained from the PDFs of the energy differences of low-energy bursts
is compatible, within the standard deviations, to that one found from earthquakes ($q=1.75\pm0.15$;
\citealt{2007PhRvE..75e5101C}). Consistent with theoretical models of FRBs based on magnetar (e.g.,
\citealt{2010vaoa.conf..129P,2014MNRAS.442L...9L,2017ApJ...843L..26B,2020ApJ...896..142B,2019MNRAS.485.4091M}),
the earthquake-like behavior here may indicate that the low-energy bursts of FRB 121102 could originate from starquakes of a magnetar.
More interestingly, the average $q$ values of energy, peak flux, and duration for the high-energy component
of FRB 121102 are roughly consistent with those of SGRs. We next show that these $q$ values can be well understood
within the same statistical framework of a SOC system.

\subsection{Predictions of the fractal-diffusive avalanche model}

For a fractal-diffusive avalanche model, quantitative values for the size distributions of SOC parameters
(length scales $L$, durations $T$, peak fluxes $P$, and fluences or energies $E$) are derived from first principles,
using the scale-free probability conjecture, $N(L)dL\propto L^{-S}dL$, for three Euclidean dimensions
$S=1$, 2, and 3. The analytical model predicts the indices of size distributions for $E$, $T$, and $P$ as
\citep{2012A&A...539A...2A,2014ApJ...782...54A}
\begin{eqnarray}\label{eq:alpha}
\alpha_{E}&=&1+(S-1)/(D_{S}+2)\nonumber\\
\alpha_{T}&=&(1+S)/2\;,\\
\alpha_{P}&=&2-1/S\nonumber
\end{eqnarray}
where $D_{S}\approx (1+S)/2$ is the mean fractal dimension.
As mentioned above, \cite{2007PhRvE..75e5101C} presented an analysis method to explain SOC behavior
in the limited number of earthquakes by making use of the return distributions (i.e., distributions of
the avalanche size differences at different times). They obtained the first strong evidence that the return
distributions appear to have the shape of $q$-Gaussians, standard distributions arising naturally in
nonextensive statistical mechanics \citep{1988JSP....52..479T,1998PhyA..261..534T}.
Under the assumption that there is no correlation between the sizes of two events, an exact relation
between the index $\alpha$ of the avalanche size distribution and the $q$ values of the appropriate
$q$-Gaussian has been obtained as \citep{2007PhRvE..75e5101C,2010PhRvE..82b1124C}
\begin{equation}
q=\frac{\alpha+2}{\alpha}\;,
\label{eq:qalpha}
\end{equation}
which is  important because it makes the $q$ parameter determined a priori from one of the well-known
indices of the system.

In the framework of the fractal-diffusive SOC model, we obtain the theoretical $q$ values, i.e.,
$q_{E}\approx2.33$, $q_{P}=2.2$, and $q_{T}=2.0$, according to Equations~(\ref{eq:alpha})
and (\ref{eq:qalpha}) by taking the three-dimensional Euclidean space $S=3$. The observed $q$ values
of SGR J1550--5418, SGR 1806--20, and SGR J1935+2154 at $1\sigma$ confidence level are in the range of
$q_{E}\approx2.12-2.90$, $q_{P}\approx2.10-2.70$, and $q_{T}\approx1.83-2.53$, which agree well with
the predictions of the fractal-diffusive SOC model.\footnote{Due to the wide range of $q$ values, they
may also be consistent with the predictions of the fractal-diffusive SOC model with the spatial dimension $S=2$.}
We also note that the derived $q_{E}$ and $q_{P}$ of the high-energy component of FRB 121102 ($q_{E}=2.21\pm0.10$
and $q_{P}=2.34\pm0.09$) are in good agreement with the predictions of the fractal-diffusive SOC model,
and the derived $q_{T}$ ($1.50\pm0.16$) is also roughly consistent with the model prediction
at $3.1\sigma$ confidence level. Therefore, SGRs and high-energy bursts of FRB 121102 share similar scale-invariant
behavior, suggesting that they can be explained within the same statistical framework of fractal-diffusive SOC systems
with the spatial dimension $S=3$.

The observed durations of the repeating bursts of FRB 121102 are directly used for the analysis of scale invariance.
As is well known, the observed pulse duration of an FRB is generally broadened by the instrumental and astrophysical
sources \citep{2003ApJ...596.1142C}. The instrumental pulse broadening
includes the sampling time-scale and the intrachannel dispersion smearing. The temporal resolution can
not be better than the sampling time-scale, which broadens the pulse. The smearing is caused by intrachannel
dispersion. The astrophysical pulse broadening is the scattering process of the radio emission
in inhomogeneous plasma. Therefore, the observed duration does not directly represent the intrinsic duration.
The intrinsic durations of the repeating bursts, in principle, should be
used for our purpose. In practice, it is difficult to separate the intrinsic duration from
the broadening components. But note that the distribution of the observed duration differences at different
time intervals is studied in this work. Pulse broadening caused by smearing and scattering can be approximately
deducted from subtracting two observed durations. That is, the difference between two observed durations
can be approximated as the difference between two intrinsic ones.

\section{Summary and Discussion}
\label{sec:summary}
Previous studies have shown that the PDFs for the earthquake energy differences at different time intervals have fat tails
with a $q$-Gaussian shape \citep{2007PhRvE..75e5101C,2015EPJB...88..206W}. Another remarkable feature is that
the $q$ values in the $q$-Gaussian distributions are approximately equal and independent of the temporal interval scale
adopted, indicating a scale-invariant structure in the energy differences of earthquakes. These statistical features
can be well explained within a dissipative SOC scenario taking into account long-range interactions \citep{2007PhRvE..75e5101C}.
That is, scale invariance is an important character representing the system approaching to a critical state.
Recently, \cite{2017ChPhC..41f5104C} found that SGR J1550--5418 also has the property of scale invariance.
However, it is unclear whether other SGRs share the same scale-invariant structure as SGR J1550--5418.

In this work, we present statistics of soft gamma-/hard X-ray bursts from SGR 1806--20 and SGR J1935+2154. We find that
the two SGRs share a common behavior in terms of the PDFs of the fluence differences, confirming
the findings of \cite{2017ChPhC..41f5104C}. The PDFs of the fluence differences can be well fitted by a $q$-Gaussian
function and the $q$ values keep nearly constant for different scale intervals. These results
support that there is a common scale-invariant structure in SGRs and that systems responsible for SGR bursts
are in a SOC state. Moreover, we show that the burst fluxes and durations of SGRs share similar
scale-invariant behaviors with fluences, i.e., the PDFs of the differences of fluxes and durations at different
time intervals also well follow $q$-Gaussian distributions.

Given the association between an FRB-like event and an X-ray burst from the Galactic magnetar SGR J1935+2154,
we also investigate the scale-invariant structure of the repeating FRB 121102. For FRB 121102, we show that
the PDFs of the differences of energies, fluxes, and durations also exhibit $q$-Gaussian distributions, with
steady $q$ values independent of the time scale. These properties are very similar to those of SGRs,
and thus both can be attributed to a SOC process. Due to the fact that the burst energy distribution
of FRB 121102 is bimodal, we also investigate the scale-invariant behaviors of low- and high-energy
bursts of FRB 121102. We find that the $q_{E}$, $q_{P}$, and $q_{T}$ values of low-energy bursts are different
from those of high-energy bursts, which indicate that the low-energy component and the high-energy component
may have different physical origins. Interestingly, the average $q$ value obtained from the PDFs of the
energy differences of low-energy bursts is close to the $q$ value found from earthquakes. This implies that
there may be some similarities between the energy origins of low-energy bursts of FRB 121102 and earthquakes.
More interestingly, the average $q$ values of energy, peak flux, and duration for the high-energy component
of FRB 121102 are roughly consistent with those of SGRs. These $q$ values can be well explained within the same
statistical framework of fractal-diffusive SOC systems with the spatial dimension $S=3$.
In the future, much more repeating bursts of FRBs will be detected. The physical connection between repeating
FRBs and SGRs can be further investigated.

\section*{Acknowledgements} %\acknowledgments
We would like to thank the anonymous referee for helpful comments.
This work is partially supported by the National Natural Science Foundation of China
(grant Nos.~11988101, 11725314, U1831122, 12041306, and U1831207), the Youth Innovation Promotion
Association (2017366), the Key Research Program of Frontier Sciences (grant No.
ZDBS-LY-7014) of Chinese Academy of Sciences, and the Major Science and Technology
Project of Qinghai Province (2019-ZJ-A10). ZGD was supported by the National Key Research
and Development Program of China (grant No. 2017YFA0402600), the National SKA Program of China
(grant No. 2020SKA0120300), and the National Natural Science Foundation of China (grant No. 11833003).

%\bibliographystyle{apj}
%\bibliography{ms}

\end{document}